\documentclass[aps,preprint]{revtex4}%
\usepackage{amsfonts}
\usepackage{amsmath}
\usepackage{amssymb}
\usepackage{graphicx}%
\begin{document}
\title{On "Inconsistency in the application of the adiabatic theorem"}
\author{Z. Wu, L. Zheng, and H. Yang}
\affiliation{Center for Theoretical physics, Jilin University, Changchun, Jilin 130023, China}
\maketitle

In a recent letter[1], Marzlin and Sanders gave a proof of inconsistency
implied by the adiabatic theorem(AT), and declared that the standard
statetment of the AT, (3) in [1], alone does not ensure that a formal
application of it result in correct results, and that the implication of MS
inconsistency is important. This has attracted attention from the physics
circle[2,3,4]. Though Mazlin and Sanders pointed out that "We show that a
perfunctory application of this statement is problematic if \ldots\ ,
regardless of \ldots", their explanation for a perfunctory use of AT is
specious. Their proof of inconsistency is a step by step formal derivation.
The only convincing way of refuting it is to point out which step of the
derivation is logically invalid. But Marzlin and Sanders failed to do this.
Instead, they just gave an explanation, which, as we will show, is wrong. In
view of the importance of the AT, the purpose of this note is to point out
where their proof goes wrong, why their explanation is specious and the MS
inconsistency is just one of the contradictions obtained from wrong reasoning
and has nothing to do with physics.

To avoid the infinitely large time variable $t$, and infinitely small
coefficients in (1) of [1], let's get to the scaled dimenssionless time
variable $\tau=t/T$, where $T$ is the large total evolution time, during which
the Hamiltonian has experienced finite change. The eq.(1) of [1] can be
written as%
\begin{equation}
i\frac{\partial}{\partial\tau}\psi_{n}=-i\sum_{m}\langle E_{n}|\frac{d}{d\tau
}E_{m}\rangle e^{i\int(E_{n}-E_{m})}\psi_{m} \label{eqn1}%
\end{equation}

"$|\psi_{AT}(t)\rangle$ is an approximate solution to the Schr\"{o}dinger
equation (1)" means only "$|\psi_{AT}(t)\rangle\approx|\psi_{EX}(t)\rangle$,
where $|\psi_{EX}(t)\rangle$ is the exact solution." Considering the basic
mathematical fact that "$|\psi_{AT}(t)\rangle\approx|\psi_{EX}(t)\rangle$ does
not imply $|\overset{\cdot}{\psi}_{AT}(t)\rangle\approx|\overset{\cdot}{\psi
}_{EX}(t)\rangle$", one easily sees that substituting the approximate solution
(3) of [1] back into the Schr\"{o}dinger equation (1) is logically invalid. It
leads directly to contradictions (see [5] for the detail),%

\begin{equation}
\langle E_{n}(T\tau)|\frac{d}{d\tau}E_{m}(T\tau)\rangle\approx0,\forall m\neq
n
\end{equation}

"Eq.(3) of [1] does not ensure that eq.(1) approximately holds" means the
following approximate equation does not hold $i\hbar\frac{\partial}%
{\partial\tau}U_{AT}(T\tau,T\tau_{0})\approx TH(T\tau)U_{AT}(T\tau,T\tau_{0}%
)$. Therefore the statement $i\hbar\frac{\partial}{\partial\tau}U_{AT}%
(T\tau,T\tau_{0})^{\dag}|E_{0}(T\tau_{0})\rangle\approx$%

\begin{equation}
-TU_{AT}(T\tau,T\tau_{0})^{\dag}H(T\tau)U_{AT}(T\tau,T\tau_{0})U_{AT}%
(T\tau,T\tau_{0})^{\dag}|E_{0}(T\tau_{0})\rangle
\end{equation}

is invalid. Combining eq.(3) and $i\hbar\frac{\partial}{\partial\tau}%
\exp\left\{  \frac{i}{\hbar}T\int E_{0}(T\tau^{\prime})d\tau^{\prime}\right\}
|E_{0}(T\tau_{0})\rangle$

$=-TU_{AT}(T\tau,T\tau_{0})^{\dag}H(T\tau)U_{AT}(T\tau,T\tau_{0})\exp\left\{
\frac{i}{\hbar}T\int E_{0}(T\tau^{\prime})d\tau^{\prime}\right\}  |E_{0}%
(T\tau_{0})\rangle,$ which holds exactly, Marzlin and Sanders proved thier
claim (4) in [1] and hence the MS inconsistency (6) in [1]. Thus we have
pointed out that one of the premises of thier proof, eq.(19), is wrong, so
their claim (6) and MS inconsistency (7) are unfounded results.

The paragraph beneath (9) in [1] is their explanation. Let's show, it is
wrong. In fact, substitute (3) of [1] back into the Schr\"{o}dinger equation
(1), means neglect of the non-diagonal matrix elements terms, which
\textbf{are rapidly oscillating}. But this neglect still leads to
contradictions (2). Expanding $|\frac{d}{d\tau}E_{m}(T\tau)\rangle$ in terms
of $|E_{n}(T\tau)\rangle$ and using (2), one gets $|\frac{d}{d\tau}E_{m}%
(T\tau)\rangle\approx|E_{m}(T\tau)\rangle\left\langle E_{m}(T\tau)|\frac
{d}{d\tau}E_{m}(\check{R}(\tau))\right\rangle $. Taking the inner products of
both sides with $|E_{m}(T\tau_{0})\rangle$, we get $\frac{d}{d\tau
}\left\langle E_{m}(T\tau_{0}))|E_{m}(T\tau)\right\rangle \approx\left\langle
E_{m}(T\tau))|\frac{d}{d\tau}E_{m}(T\tau)\right\rangle \left\langle
E_{m}(T\tau_{0})|E_{m}(T\tau)\right\rangle $. Solving this differential
equation gives the MS inconsistency $\left\langle E_{m}(T\tau_{0})|E_{m}%
(T\tau)\right\rangle \approx\exp\{\int_{\tau_{0}}^{\tau}\left\langle
E_{m}(T\tau^{\prime})|\frac{d}{d\tau}E_{m}(T\tau^{\prime})\right\rangle
d\tau^{\prime}\}$.

Thus we have shown that the explanation of [1] is wrong, and MS inconsistency
is just one of the contradictions from wrong reasoning, it has nothing to do
with physics and is not worth noting at all.

\end{document}